\documentstyle[aps,epsfig,prl,floats]{revtex}
\def\NdCe{Nd$_{2-x}$Ce$_{x}$CuO$_4$ }
\def\cm-1{cm$^{-1}$}
\def\Tc{$T_{c}$ }
\begin{document}
\draft

\title{Nonmonotonic $d_{x^2-y^2}$ Superconducting Order Parameter in \NdCe}

\author{
G.~Blumberg$^{1,\dag}$, A.~Koitzsch$^{1}$, A.~Gozar$^{1}$, 
B.S.~Dennis$^{1}$,  C.A.~Kendziora$^{2}$, 
P.~Fournier$^{3,\P}$, and R.L.~Greene$^{3}$
}
\address{
$^{1}$Bell Laboratories, Lucent Technologies, Murray Hill, New Jersey 
07974 \\
$^{2}$United States Naval Research Laboratory, Code 6333, Washington, 
D.C. 20375 \\
$^{3}$Center for Superconductivity Research and Department of Physics, 
University of Maryland, College Park, MD 20742 \\
}

\twocolumn[\hsize\textwidth\columnwidth\hsize\csname
@twocolumnfalse\endcsname
\date{June 15, 2001; accepted for PRL}
\maketitle
\widetext

\begin{abstract}
    Low energy polarized electronic Raman scattering  
    of the electron doped superconductor \NdCe ($x=0.15$, $T_{c}=22$~K) 
    has revealed a \emph{nonmonotonic} $d_{x^2-y^2}$ superconducting 
    order parameter.  
    It has a maximum gap of $4.4k_{B}T_{c}$ at Fermi surface intersections 
    with antiferromagnetic Brillouin zone (the ``\emph{hot spots}'') and 
    a smaller gap of $3.3k_{B}T_{c}$ at fermionic Brillouin zone boundaries. 
    The gap enhancement in the vicinity of the ``\emph{hot spots}'' 
    emphasizes role of antiferromagnetic fluctuations and similarity in 
    the origin of superconductivity for electron- and hole-doped cuprates. 
\end{abstract}

]

\narrowtext

{\em Introduction}.--
\NdCe (NCCO) is one of a few electron doped cuprate superconductors 
\cite{Tokura:89}.  
The physical properties of the electron doped cuprates are different 
from the hole-doped. 
Structurally, NCCO does not have apical oxygen atoms. 
It is commonly believed that the charge carriers in NCCO are electrons 
rather than holes as in other cuprate families \cite{Tokura:89,twoband}. 
In optimally hole doped cuprates the normal state resistivity 
increases linearly over a wide range of temperatures while for NCCO  
the in-plane resistivity is quadratic in temperature with a large 
residual value \cite{Peng:97}. 
For the electron doped cuprates the superconducting (SC) transition 
temperature is relatively low and the 
superconductivity occurs in a narrow doping range \cite{Tokura:89}. 
From the early tunneling \cite{Huang:90} and microwave measurements an 
$s$-wave SC order parameter (OP) was suggested \cite{Wu:93} 
that is in contrast with the $d$-wave symmetry established for hole 
doped compounds. 
The early Raman measurements were interpreted as  
evidence for nearly uniformly gapped Fermi surface (FS) 
\cite{Stadlober:95} 
consistent with the $s$-wave OP. 
However, the interpretation of more recent
microwave measurements \cite{Kokales:00,Prozorov:00} along with
angle resolved photoemission spectroscopy (ARPES) 
\cite{Armitage:01,Sato:01}, and phase sensitive scanning SQUID 
microscope experiments \cite{dwave} are consistent with $d$-wave OP. 

We report polarized low energy electronic Raman scattering studies on NCCO 
single crystals and show that the data is consistent with a SC OP of 
the  $d_{x^2-y^2}$ symmetry. 
However, as distinguished from the simplest commonly assumed SC gap  
function, $\Delta({\bf{k}}) \propto \cos({k_{x}a}) - \cos({k_{y}a)}$,  
where $\bf{k}$ is a wave vector on the FS and $a$ is the 
$ab$-plane lattice constant, the present results require a 
\emph{nonmonotonic} form of the OP. 
We find that in contrast with hole doped cuprates for NCCO the positions 
of the SC gap maxima are located closer 
to the nodes than to the Brillouin zone (BZ) boundaries.  
The gap opens up rapidly with departure from the diagonal nodal directions 
and quickly reaches its maximum value of $4.4k_{B}T_{c}$ at the 
intersections of the FS and the antiferromagnetic (AF) BZ. 
However, the gap value drops to $3.3k_{B}T_{c}$ at the BZ boundaries. 
The implications of such \emph{nonmonotonic} OP to the doping dependence of 
\Tc are discussed.

{\em Experimental}.--
The Raman experiments were performed from a natural $ab$ surface of 
a plate-like single crystal grown as 
described in Ref. \cite{Peng:91}. 
After growth, the crystal was annealed in an oxygen-reduced 
atmosphere to induce the doping level for optimal \Tc. 
SC transition measured by SQUID was about 22~K with a 
width about 2~K. 
The sample was mounted in an optical continuous helium flow cryostat. 
Spectra were taken in a backscattering geometry using linearly
polarized excitations of a Kr$^{+}$ laser from near infrared to violet.
An incident laser power less than 2~mW was focused to a 50~$\mu$m
spot onto the sample surface.
The referred temperatures were corrected for laser heating. 
The spectra were measured at temperatures between 5 and 35~K and 
were analyzed by a custom triple grating spectrometer. 
The data were corrected
for the spectral response of the spectrometer and for the optical
properties of the material at different wavelengths as described in
the Ref.~\cite{Blumberg:94}.  

The polarization directions of the incident ($\textbf{e}_{i}$) and 
scattered ($\textbf{e}_{s}$) photons are indicated by 
($\textbf{e}_{i}, \textbf{e}_{s}$) with $x = [100]$, $y = [010]$, 
$x^{\prime} = [110]$, and $y^{\prime} = [\bar{1}10]$. 
The presented data were taken in ($xy$), ($x^{\prime}y^{\prime}$), and 
($xx$) scattering geometries. 
For tetragonal $D_{4h}$ symmetry these geometries correspond to 
spectra of $B_{2g} + A_{2g}$, $B_{1g} + A_{2g}$, and $A_{1g} + B_{1g}$ 
representations. 
In addition, by using geometries with circularly polarized light we  
checked the intensity of the $A_{2g}$ component and found it to be 
negligibly weak. 

{\em Raman scattering symmetries}.--
  \begin{figure}[t]
      \begin{center}
      \epsfig{file=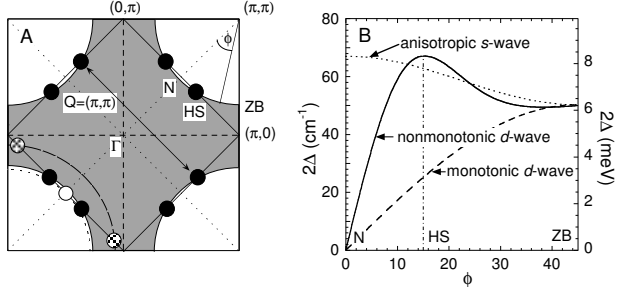, width=8.5cm}
      \caption[]
             {
	     (A) A schematic representation of the electron doped FS of NCCO 
	     \cite{Armitage:01}. 
	     The occupied electron states are shaded.  
	     The AF BZ at half filling is shown as the square rotated by 
	     $45^{\circ}$. 
	     AF fluctuations enhance interactions between fermions 
	     around the ``hot spots'' (filled circles), 
	     the regions of the FS connected by the 
	     $\textbf{Q} = (\pi, \pi)$ vector \cite{Rice:95}.
	     The location of the ``hot spots'' sensitively depend on 
	     the doping level. 
	     The FS shrinks with further electron doping until the 
	     intersection with the AF BZ vanishes (dotted lines and  
	     empty circle in the lower left quadrant). 
	     The hole doped cuprates exhibit a large FS with the ``hot spots'' 
	     shifted to the vicinity of $(\pi, 0)$ and the equivalent 
	     points (dashed lines and hatched circles in the lower 
	     left quadrant) \cite{Norman:98}.   
	     The dotted diagonal (dashed horizontal and vertical) lines denote 
	     the nodes of the $B_{1g}$ ($B_{2g}$) Raman form factor. 
	     (B) The magnitude of the $d_{x^2-y^2}$ OP as a function of the 
	     angle $\phi$ along the FS. 
	     Solid line: {\em non-monotonic} OP for NCCO.
	     The gap value rises rapidly from the nodal diagonal direction 
	     (N) to its maximum value $2\Delta = 67$~\cm-1 at the ``hot 
	     spot'' (HS) observed in the $B_{2g}$ channel.  
	     The $2\Delta$-peak at 50~\cm-1 in the $B_{1g}$ channel 
	     corresponds to the value at the BZ boundary (ZB). 
	     Dashed line: {\em monotonic} $\sin(2 \phi)$ form. 
	     Dotted line: anisotropic $s$-wave OP proposed in the 
	     Ref.~\cite{Stadlober:95}. 
	     }
      \label{Schematics}
      \end{center}
  \end{figure}
The electronic Raman response function for a given geometry ($\textbf{e}_{i}, 
\textbf{e}_{s}$) is proportional to the sum over the density of states 
at the FS weighted by the momentum $\textbf{k}$ dependent form 
factor \cite{Dierker,Miles,Einzel}. 
By choosing the scattering geometries one can selectively probe 
different regions of the FS and obtain information about 
the $\textbf{k}$ dependence of the SC OP. 
For the $B_{1g}$ channel the Raman spectrum has a form
factor of $d_{x^{2}-y^{2}}$ symmetry that vanishes at 
the $(0, 0) \rightarrow (\pi, \pi)$ and the equivalent 
diagonal lines of the BZ (See Fig.~\ref{Schematics}). 
The spectrum intensity in the $B_{1g}$ channel integrates mainly from 
the regions of the FS distant from these diagonals, near 
intersections of the FS and the BZ boundary (ZB). 
In contrast, the form factor for $B_{2g}$ spectrum has $d_{xy}$ 
symmetry and therefore vanishes along $(0, 0) \rightarrow (0, \pi)$ 
and the equivalent lines. 
The intensity in the $B_{2g}$ channel is mainly determined by  
excitations near $(\pi/2, \pi/2)$ and the equivalent points. 
All regions of momentum space may contribute to the fully symmetric 
$A_{1g}$ channel \cite{A1g}. 

{\em The pair breaking excitations}.--
  \begin{figure}[t]
      \begin{center}
      \epsfig{file=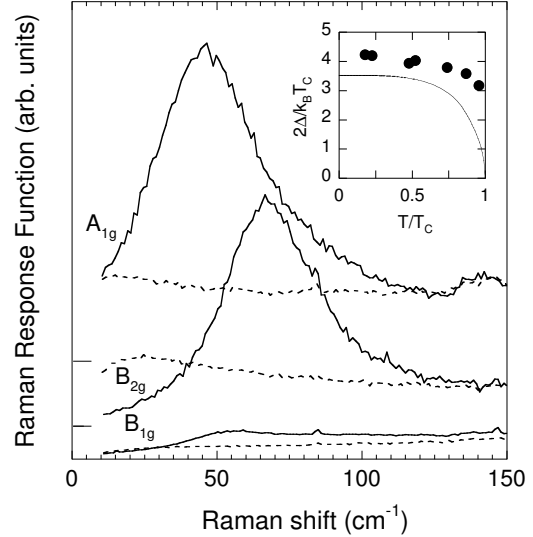, width=7.0cm}
      \vspace{3mm}
      \caption[]
             {Low-frequency Raman scattering spectra with 1.9~eV 
             excitation for different symmetry channels. 
	     The solid lines denote spectra at 11~K in the SC  
	     state and the dashed lines spectra taken above \Tc at 35~K.  
	     The baselines are shifted as indicated by the ticks. 
	     The $B_{1g}$ spectra corresponds to $x^{\prime}y^{\prime}$, 
	     the $B_{2g}$ to $xy$ and the $A_{1g}$ to 
	     $xx - x^{\prime}y^{\prime}$ scattering geometries. 
	     The inset shows the temperature evolution of the $2\Delta$-peak 
	     energy in the $B_{2g}$ channel. 
	     The line indicates the mean field BCS temperature dependence. 
	     }
      \label{Two-delta}
      \end{center}
  \end{figure}
In the Fig.~\ref{Two-delta} we compare the low energy Raman spectra 
above and below the SC transition taken with the excitation energy 
$\omega_{L} =1.9$~eV for three symmetry representations: $B_{1g}$, 
$B_{2g}$ and $A_{1g}$. 
Above \Tc spectra exhibit a flat electronic Raman continuum. 
In the SC state the low-frequency tail of the Raman continuum changes to 
reflect the opening of the SC gap:  
the strength of the low-frequency continuum is reduced and the spectrum  
acquires the so-called $2\Delta$-peak as a result of excitations 
across the anisotropic gap, $2\Delta(\bf{k})$.
These peaks correspond to the excitations out of the SC 
condensate.
For different scattering geometries spectra differ in their intensity 
as well as in the position of the $2\Delta$-peaks. 
The peaks in the $A_{1g}$ and $B_{2g}$ channels are an 
order of magnitude stronger than in the $B_{1g}$ channel.
For $B_{2g}$ symmetry the peak is at the highest energy, at about  
67~\cm-1,  followed by the peaks in $B_{1g}$ and $A_{1g}$ channels at 50 
and 40~\cm-1 correspondingly. 

These results are in sharp contrast to the hole doped 
cuprates where the most prominent scattering is observed in $B_{1g}$ 
channel for which the $2\Delta$-peak is at the highest frequency  
\cite{Lance-Ygap,Staufer,Chen-La,Devereaux:94,Kang:96,Blumberg:97,Liu:99}.  
For the hole doped cuprates the interpretation of Raman data is consistent 
with $d_{x^2-y^2}$ SC OP of the simplest {\em monotonic} $\propto 
\sin(2 \phi)$ form shown in the Fig.~\ref{Schematics}B. 
The role of orthorhombic distortions and impurities has been discussed 
in the Refs. \cite{Einzel,Beal-Monod:97,Nemetschek:98,Devereaux:97}.  
 
\begin{figure}[t]
      \begin{center}
      \epsfig{file=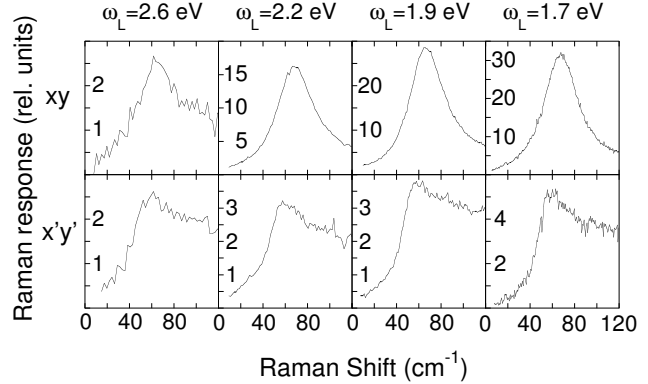, width=8.5cm}
      \vspace{3mm}
      \caption[]
             {Low energy Raman spectra at 8~K in the $B_{2g}$ ($xy$) and 
             $B_{1g}$ ($x^{\prime}y^{\prime}$) channels 
             for excitations from blue to near IR. 
	     Note that for shorter wavelength excitations the intensities of 
	     the $2\Delta$-peaks in two channels are comparable while for the 
	     red excitations the intensities in the $B_{2g}$ channel 
	     is an order of magnitude stronger than in the $B_{1g}$ channel. 
	     }
      \label{Excitations}
      \end{center}
  \end{figure}

The earlier low-temperature Raman data from NCCO was measured down to 
about 25~\cm-1 with $\omega_{L} =2.6$~eV \cite{Stadlober:95}.
The data exhibited a strong residual scattering intensity.    
The observed $2\Delta$-peak in the $B_{1g}$ channel started at a threshold 
from the low energy side. 
The authors discuss possible experimental artifacts for the residual 
intensity and suggest that the observed threshold 
supports an anisotropic $s$-wave gap interpretation (see 
Fig.~\ref{Schematics}B). 
Our data extends to much lower frequencies  
(Figs.~\ref{Two-delta}-\ref{Excitations}).  
The spectra for all scattering channels show a smoothly dropping intensity 
below the $2\Delta$-peak down to the lowest energies measured. 
Based on our data we exclude the anisotropic $s$-wave interpretation since 
any fully gapped FS would lead to a Raman intensity threshold 
as it has been observed for classical superconductors \cite{Dierker}.   
The smooth decrease of the scattering intensity is a signature of the nodes 
in the OP \cite{Devereaux:94}. 

{\em The gap anisotropy}.--
The observation of the $2\Delta$-peak in the $B_{2g}$ channel at 
energies higher than in the $B_{1g}$ channel suggests a {\em 
nonmonotonic} OP with maxima in $\Delta(\bf{k})$ closer to the $(0, 0) 
\rightarrow (\pi, \pi)$ diagonal than to the BZ boundary. 
The recent ARPES studies of NCCO exhibit a node in $\Delta(\bf{k})$  
along this diagonal direction \cite{Armitage:01,Sato:01}. 
Our Raman data can be reconciled with the ARPES results by including 
higher harmonics, like $\sin(6 \phi)$, to the {\em monotonic} $\sin(2 
\phi)$ form of the $d_{x^2-y^2}$ OP. 
The resulting {\em nonmonotonic} form is shown in Fig.~\ref{Schematics}B. 
In Fig.~\ref{Schematics}A we sketch the FS  
as seen by ARPES \cite{Armitage:01}. 
The latter data  exhibits regions of suppressed 
spectral weight at the intersections with the AF BZ 
boundary, a behavior  
similar to the destruction of the FS at the ``hot spots'' in the pseudogap 
phase seen in hole-doped cuprates \cite{Norman:98,Shen:98}. 
Strong AF fluctuations are believed to be responsible for such ``hot 
spot'' behavior \cite{Rice:95,Chubukov:97,Schmalian:98,Ioffe:98}. 

We assume that the AF interactions are responsible for the 
SC coupling mechanism and that like in the hole-doped cuprates the SC gap 
reaches its maximum value in the vicinity of the ``hot spots''. 
For the hole-doped cuprates with a large FS the ``hot spots'' are close 
to the BZ boundary.  
For electron-doped cuprates the position of the ``hot spots'' 
sensitively depends on the size of the FS and, hence, on the amount of 
doping.   
As it is seen in the ARPES data \cite{Armitage:01} for the optimally 
doped NCCO the ``hot spots'' are close to the BZ diagonals and 
therefore in Raman the maximum gap value appears in the $B_{2g}$ 
channel. 
The peak position at about 67 \cm-1 is consistent with the maximum gap 
value of 3.7~meV observed in tunneling spectroscopy \cite{Huang:90}. 
The $2\Delta$-peak in the $B_{1g}$ channel reflects the gap 
magnitude at the BZ boundary (See Fig.~\ref{Schematics}B). 
Indeed, the peak position at about $2\Delta_{B_{1g}} = 50$~\cm-1 corresponds 
to 3~meV for a single $\Delta({\bf k}_{ZB})$ that is consistent with the 
leading edge gap estimates between 1.5 and 3~meV at the BZ boundary by ARPES 
\cite{Armitage:01,Sato:01}. 
For slightly stronger electron doping the intersection of the AF BZ and 
the FS disappears.
This naturally explains the narrow doping range for 
superconductivity in the electron-doped cuprates. 

{\em The superconducting gap temperature dependence}.--
In the inset of Fig.~\ref{Two-delta} we show the temperature development 
of the $2\Delta$-peak position in the $B_{2g}$ channel. 
The SC gap opens up very rapidly with cooling below \Tc and 
soon approaches its maximum value $4.4k_{B}T_{c}$ which 
is within the margin of the strong coupling limit and is close to the 
gap value observed for heavily hole overdoped cuprates.  
Optimally and especially hole underdoped cuprates exhibit much 
larger gap values \cite{Blumberg:97,Liu:99}.  

{\em The resonant Raman excitation profile}.--
We performed a systematic study of the Raman scattering 
efficiency as a function of the excitation photon energy.  
The low frequency response at 8~K in the $B_{1g}$ and $B_{2g}$ channels 
for excitations from blue to near IR are shown in Fig.~\ref{Excitations}. 
For the blue excitation ($\omega_{L} = 2.6$~eV) our data is consistent 
with the earlier results of Ref. \cite{Stadlober:95} showing 
comparable intensities in both $B_{1g}$ and $B_{2g}$ channels. 
The relative intensities change drastically when the excitation energy 
is decreased below 2.5~eV. 
While the peak in the $B_{1g}$ channel only slightly increases in intensity 
the peak in the $B_{2g}$ channel rapidly increases by an order of 
magnitude and exhibits a maximum around excitation $\omega_{L} = 1.9$~eV.  
  \begin{figure}[t]
      \begin{center}
      \epsfig{file=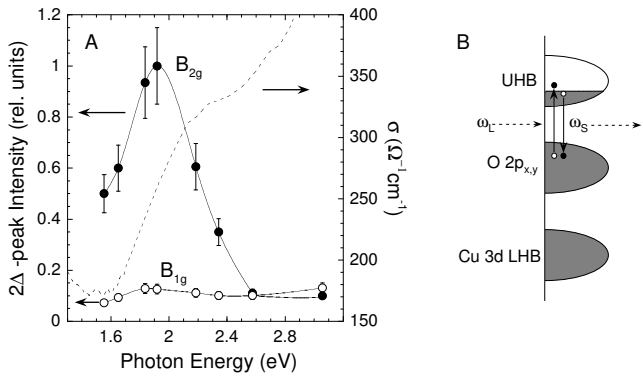, width=8.5cm}
      \vspace{3mm}
      \caption[]
             {(A) Intensity of the $2\Delta$-peaks at 8~K in the 
             $B_{1g}$ ($\circ$) and $B_{2g}$ ($\bullet$) channels as a 
             function of the excitation photon energy $\omega_{L}$ compared 
             to the optical conductivity data (dashed line) at 300~K 
             \cite{Singley:01}.   
	     The solid lines are guides to the eye. 
	     (B) A schematic diagram for the resonant electronic Raman 
             scattering process in electron doped cuprates. 
	     }
      \label{RREP}
      \end{center}
  \end{figure}
The resonance profiles of the $2\Delta$-peak 
for both $B_{1g}$ and $B_{2g}$ channels are presented in the 
Fig.~\ref{RREP}A and are compared with optical conductivity data 
\cite{Singley:01} that exhibits a band between 1.7 eV and 2.5~eV. 
This band has been ascribed to the charge-transfer process between the fully 
occupied oxygen $2p$ band and the upper Hubbard band (UHB) 
\cite{Uchida:91} that has been suggested to be a doubly occupied 
hybridized oxygen $2p$ and copper $3d$ state.

In the Fig.~\ref{RREP}B we show a schematic diagram for the resonant 
Raman scattering process in strongly correlated electron doped cuprates. 
The lower Hubbard band (LHB) and the oxygen band above are fully occupied.   
Doped electrons shift the Fermi energy to the UHB. 
Resonant enhancement of the Raman scattering process occurs when the energy 
of the incoming or scattered photons, or both, are in resonance with 
the interband transitions. 
Our results imply that the intermediate state for the Raman process 
is the same state which is seen near 2.1~eV in the optical conductivity. 
Moreover, because the observed resonance enhancement for the $B_{2g}$ 
channel is much stronger than for the $B_{1g}$ channel we anticipate that 
the interband transition occurs near $(\pi/2, \pi/2)$ point. 
Angle-resolved valence-band photoemission spectra indeed exhibits a 
band that is peaked at this point at about 2.6~eV below the 
Fermi energy \cite{Sato:01}. 
The excitation out of this band to the UHB  
may originate the 2.1~eV transition in the optical conductivity. 
A slight down scale in energy reflects the final state interactions. 

{\em Summary}.-- 
Low energy electronic Raman scattering has been investigated for 
NCCO single crystals and  
the role of interband transitions for the resonant Raman coupling has 
been discussed. 
The low temperature data for different scattering channels has been 
found to be consistent with a \emph{nonmonotonic} functional form 
of the $d_{x^2-y^2}$ OP.  
The SC gap opens up rapidly with departure from the nodal directions and 
reaches its maximum value of $4.4k_{B}T_{c}$ at the ``hot spots'' that 
are located closer to the nodes than to the BZ boundaries where the 
gap value drops to $3.3k_{B}T_{c}$. 
The enhancement of the gap value in the proximity of the ``hot 
spots'' emphasizes the role of AF fluctuations for the 
superconductivity in the electron doped cuprates. 
Despite the strong differences between the electron- and the 
hole-doped cuprates, their superconductivity appears to share the 
same symmetry and a similar origin.

{\em Acknowledgements}.--
We acknowledge discussions with N.P.~Armitage, A.~Chubukov,  
E.J.~Singley and T.~Takahashi. 
AK is supported in part by the Studienstiftung des Deutschen
Volkes.
NRL support is provided by ONR. 
UM support is provided by NSF contract DMR-9732736. 
P.F. acknowledges the support from CIAR, NSERC and the Foundation 
FORCE (Sherbrooke).

\end{document}